\documentclass{moriond}

\def\be{\begin{equation}}
\def\ee{\end{equation}}
\def\bea{\begin{eqnarray}}
\def\eea{\end{eqnarray}}

\newcommand{\Photo}{}

\begin{document}
\vspace*{4cm}
\title{QCD AT HIGH-LUMINOSITY HADRON 
COLLIDERS~\footnote{Talk at the 51st Rencontres de Moriond, La Thuile, March 2016.}}

\author{F Hautmann}

\address{Rutherford Appleton Laboratory and  Theoretical  
Physics  Dept., University of Oxford 
}

\maketitle\abstracts{
This talk gives a brief  introduction 
to open questions   in 
  jet physics and QCD  which 
come to the fore 
in the high-luminosity 
regime  characterizing 
 the upcoming phase 
of the Large Hadron Collider  and      
  future   hadron colliders.}

Precision studies  of the Standard Model (SM) and searches for rare 
processes beyond the SM    in experiments  at the Large Hadron Collider (LHC)   
  rely  on the reduction of statistical errors which will be  achieved by the  increase in 
 luminosity at the LHC Run II  and further boosted by 
  the high-luminosity upgrade    foreseen for 2020. 
 
For large  luminosity, 
 kinematic regions
 characterized by small event rates 
can  be explored, while  new 
experimental challenges are posed 
by ``pile-up", i.e., the 
high number of overlaid 
proton-proton collisions. 

The focus  of this article is on 
QCD theoretical issues 
arising in  regions accessible at  high luminosity,  
and open questions  in handling high 
pile-up,  
in particular regarding the 
treatment 
of QCD jet correlations.

\begin{figure}[htbp]
\begin{center}
\includegraphics[scale=0.4]{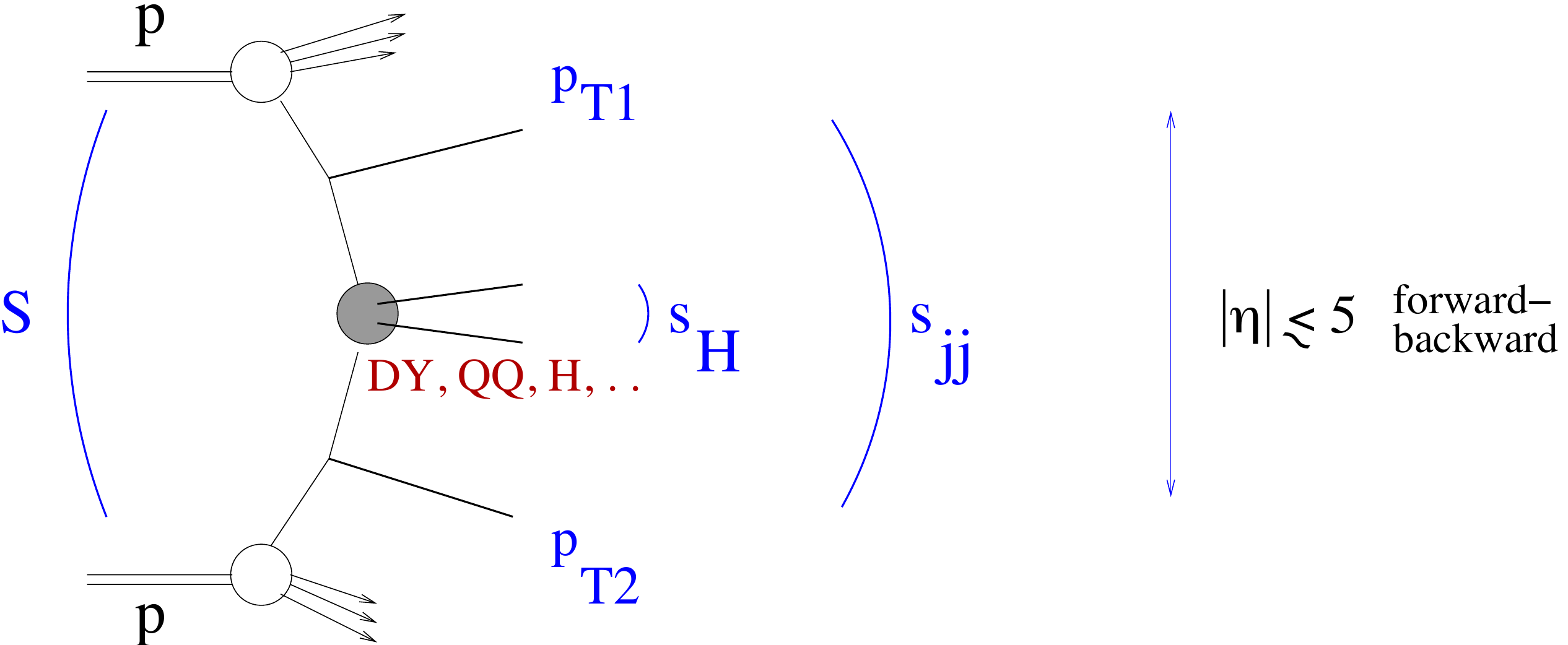}
\caption{\it   Production of  
 heavy-mass states 
(Drell-Yan lepton pairs, heavy flavor pairs, Higgs bosons, new BSM states)    
in association with  jets at the LHC.}  
\label{fig:sjj}
\end{center}
\end{figure}

Consider  the  class of 
processes depicted in 
Fig.~\ref{fig:sjj}, where 
  massive 
states such as Drell-Yan lepton pairs, 
heavy flavor pairs, Higgs bosons, or new beyond-Standard-Model  (BSM)  states    
 are produced in association 
with jets.  
LHC 
 kinematics is 
 characterized by a very 
wide range in rapidity which can be 
covered by detectors and where  
interesting physics signals can 
 appear, and by a very large range in 
the ratios of the energy scales    
$S$, $ s_{jj}$, and $s_H$   
in  Fig.~\ref{fig:sjj}.  

The  exclusive 
phase space boundary 
region $ s_H / S \to 1$ 
can  be accessed with 
high  luminosity, and 
will be  relevant to  
 searches for  
high-mass BSM states 
as well as 
SM studies such as high-mass 
$t {\bar t}$ and 
Drell-Yan production. 
This is a region where color 
radiation is strongly suppressed.     
Reliable theoretical predictions 
will  require QCD calculations 
  to take into account 
infrared gluon effects     
 from the 
imbalance between virtual processes and real emission    
to all orders in $\alpha_s$,  
and model 
nonperturbative 
color flow and color 
reconnection effects.   
Besides the 
partonic channels,  initial-state 
color-singlet channels become 
non-negligible  near the exclusive 
phase space boundary.  An example 
is  photon-photon  production,  
whose impact on BSM heavy neutral 
$Z^\prime$-boson searches 
is   analyzed in  Refs~\cite{Accomando:2016tah,Bourilkov:2016qum}.

The region 
 $ s_{ j j } \approx S   $, with 
$ s_{ j j }  \gg  s_H \gg 
\Lambda_{\rm{QCD}} $, 
is also 
relevant to  studies of 
new physics 
effects at the highest energy 
scales. Because of the large 
dijet sub-energy  this region 
probes the dynamics of  
structure functions  
and   parton cascades 
near the kinematic limit  $ x \to   1$~\cite{Hautmann:2000cq}.~However,  
since  
$ s_H /  s_{ j j }  \ll 1 $,    the phase 
space opens up for 
 finite-angle,   
multi-gluon radiation     
associated with 
scattering in 
the high-energy limit~\cite{Hautmann:2007gw}. 
The interplay of these effects has 
never been investigated  
  before and will 
 be explored  in 
high-luminosity experiments.

\begin{figure}[htbp]
\begin{center}
\includegraphics[scale=0.38]{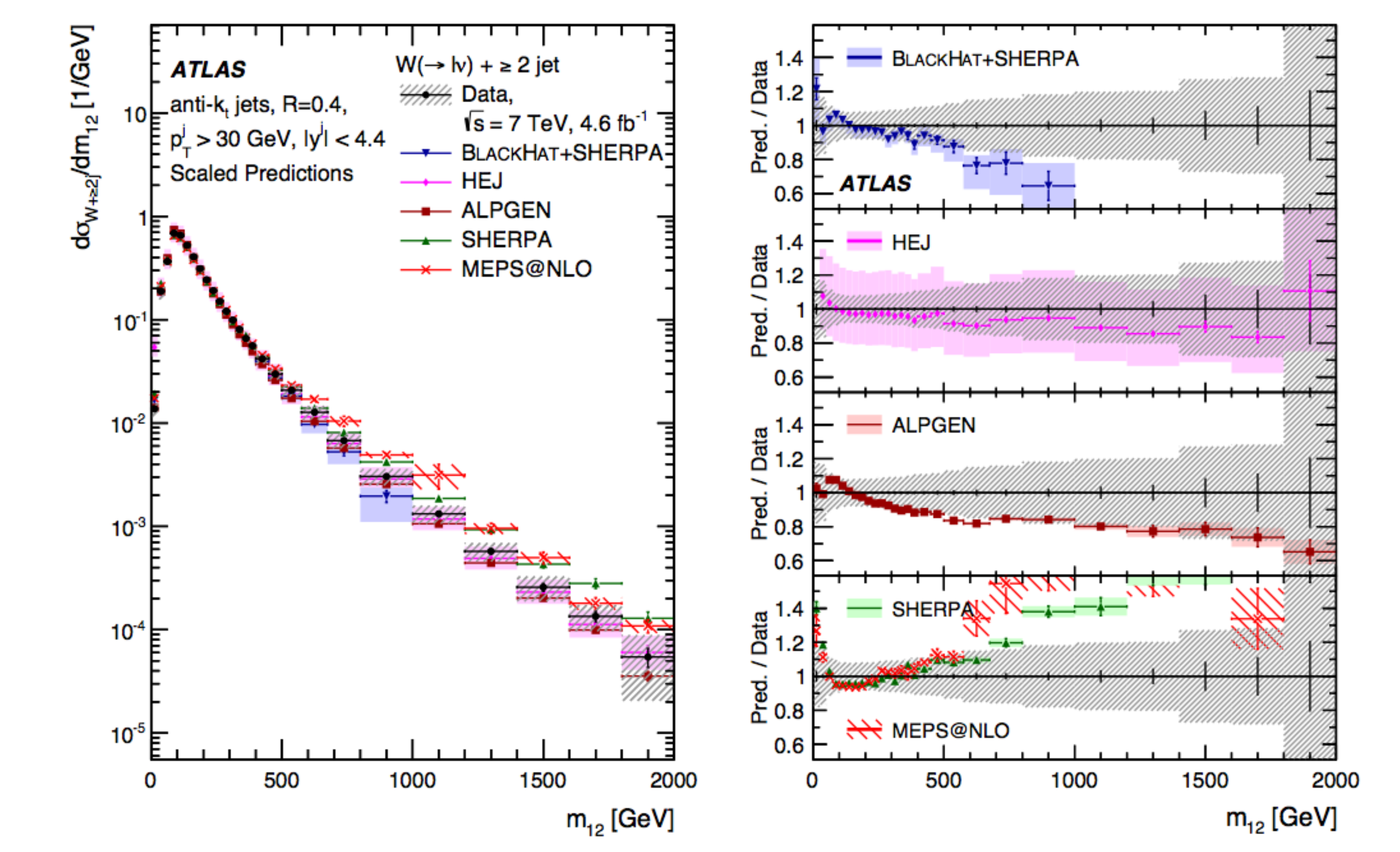}
\caption{\it Di-jet  invariant mass  measured~\protect\cite{Aad:2014qxa} 
in LHC  final states   
with $W $-boson + $2$  jets, compared 
with  parton-shower Monte Carlo calculations.}
\label{fig:REFatlasm12}
\end{center}
\end{figure}

Analyses performed at Run I have already pointed to the 
importance  of 
dijet distributions 
for increasing dijet masses ---  see 
e.g. the comparison in  
Fig.~\ref{fig:REFatlasm12}~\cite{Aad:2014qxa}  of  dijet mass measurements in 
  $W$-boson + jets   production 
with 
the  next-to-leading-order  (NLO)   
 calculation matched to parton 
shower   {\sc Blackhat}   
 + {\sc Sherpa}~\cite{Bern:2013gka}   
for masses in the region  
  above 500 GeV.   
Similar effects are seen in  Fig.~\ref{fig:REFatlasm12} from 
 the comparison with the 
{\sc Alpgen}~\cite{Mangano:2002ea} Monte Carlo.  See 
Ref.~\cite{Angeles-Martinez:2015sea} for discussion of  mass distributions 
in the context of  
transverse-momentum dependent (TMD) 
approaches to parton cascades'  evolution, which  go beyond the 
NLO+shower approach~\footnote{If TMD contributions are non-negligible 
 for large jet masses,  then extensions 
of  parton distributions  as e.g. in Refs~\cite{Hautmann:2007cx}  
 are likely to become  relevant, 
particularly  in the forward region~\cite{Deak:2009xt}.}  
 by incorporating 
 kinematical corrections due to 
energy-momentum conservation 
  constraints~\cite{Dooling:2012uw}  
and dynamical   
contributions due to finite-angle  
multi-gluon   radiation~\cite{Dooling:2014kia}.   
See Ref.~\cite{Khachatryan:2015ira} 
for  related  comparisons  of  vector  boson  
transverse momentum $p_T$ 
measurements  
in $Z$-boson + jets   production 
at Run~I 
with {\sc Madgraph}~\cite{Alwall:2011uj} 
and {\sc  Sherpa}~\cite{Gleisberg:2008ta} 
Monte Carlo calculations 
in the region of $p_T$ of 
several hundred GeV. 

With increasing luminosity, measurements of processes such as 
those in Fig.~\ref{fig:sjj}  face the challenge of 
 pile-up, which reaches 
 the level of 
50 proton-proton  collisions per 
bunch crossing at Run~II, and will be 
higher in higher-luminosity 
runs~\cite{ATLAS:2014cva}$^-$\cite{Fartoukh:2014nga}.    
  This is  especially severe in regions 
outside tracker acceptances,  where 
techniques based on  precise  
vertex and track 
reconstruction are not available.  

Different effects of pile-up 
in $Z$-boson + jets production 
are discussed   in Ref.~\cite{Hautmann:2015rrb}.  
The discussion can 
 be  extended to
other processes in Fig.~\ref{fig:sjj}. 
One effect 
consists of additional pile-up 
particles in the jet cone, leading 
to  a jet pedestal. Another  is the 
overlapping of soft particles from 
pile-up, which are clustered into 
jets. A further effect is the 
mistagging of high transverse 
momentum jets produced from 
 independent pile-up events. 
Several methods exist 
to take the first two effects into 
account and correct for them. 
These include 
techniques based on the jet vertex 
fraction~\cite{TheATLAScollaboration:2013pia} 
and charged hadron  subtraction~\cite{CMS:2014ata,Kirschenmann:2014dla},  
the {\sc Puppi} method~\cite{Bertolini:2014bba}, 
the SoftKiller method~\cite{Cacciari:2014gra}, the 
jet cleansing method~\cite{Krohn:2013lba}. 
It is pointed out in 
Ref.~\cite{Hautmann:2015yya} that the third effect is significant especially in correlation variables and can be treated using a jet mixing method. 

\begin{figure}[htb]
  \begin{center}
\includegraphics[scale=0.38, trim=0cm 5.9cm 0cm 0cm, clip=true]{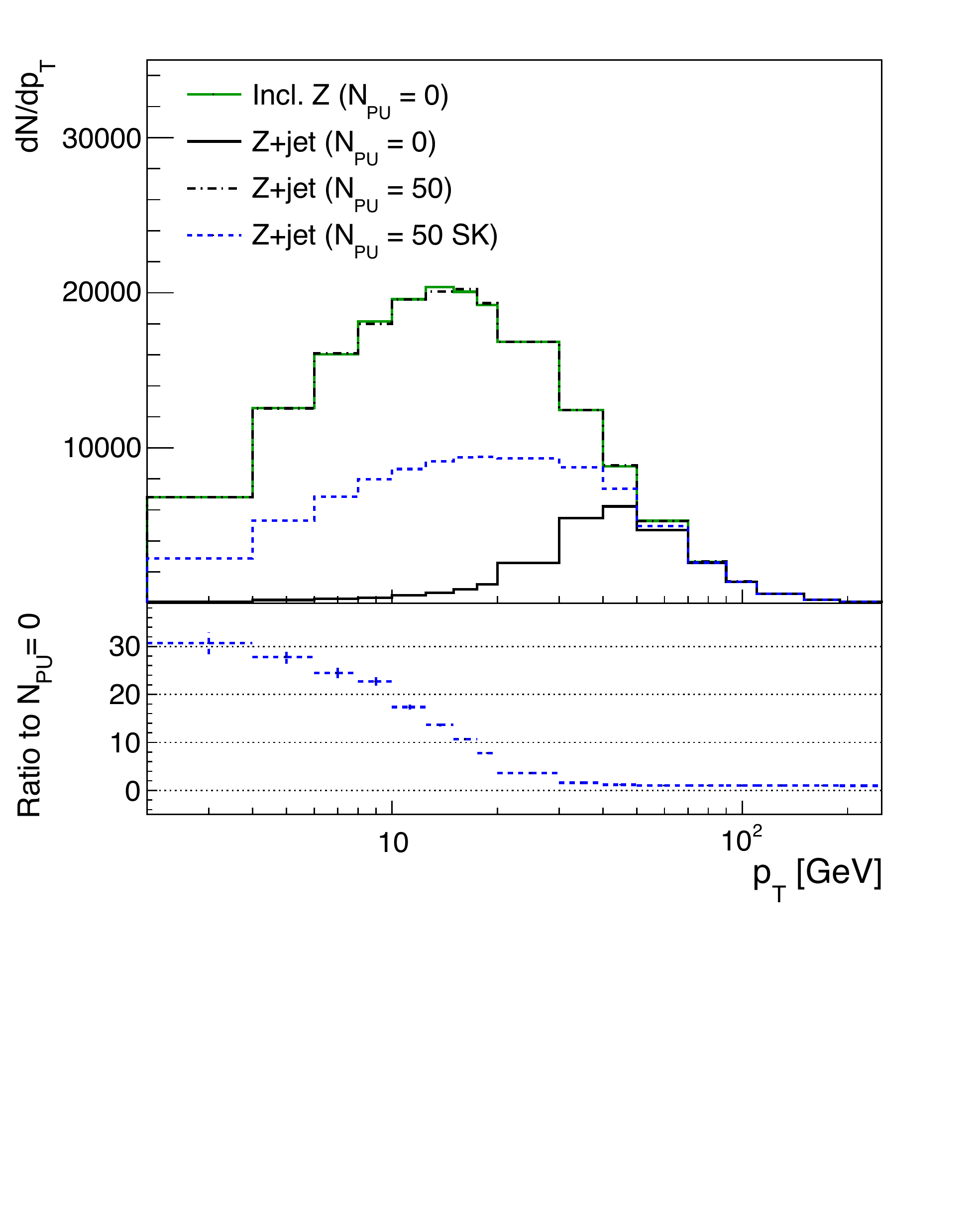}
\includegraphics[scale=0.38, trim=0cm 5.9cm 0cm 0cm, clip=true]{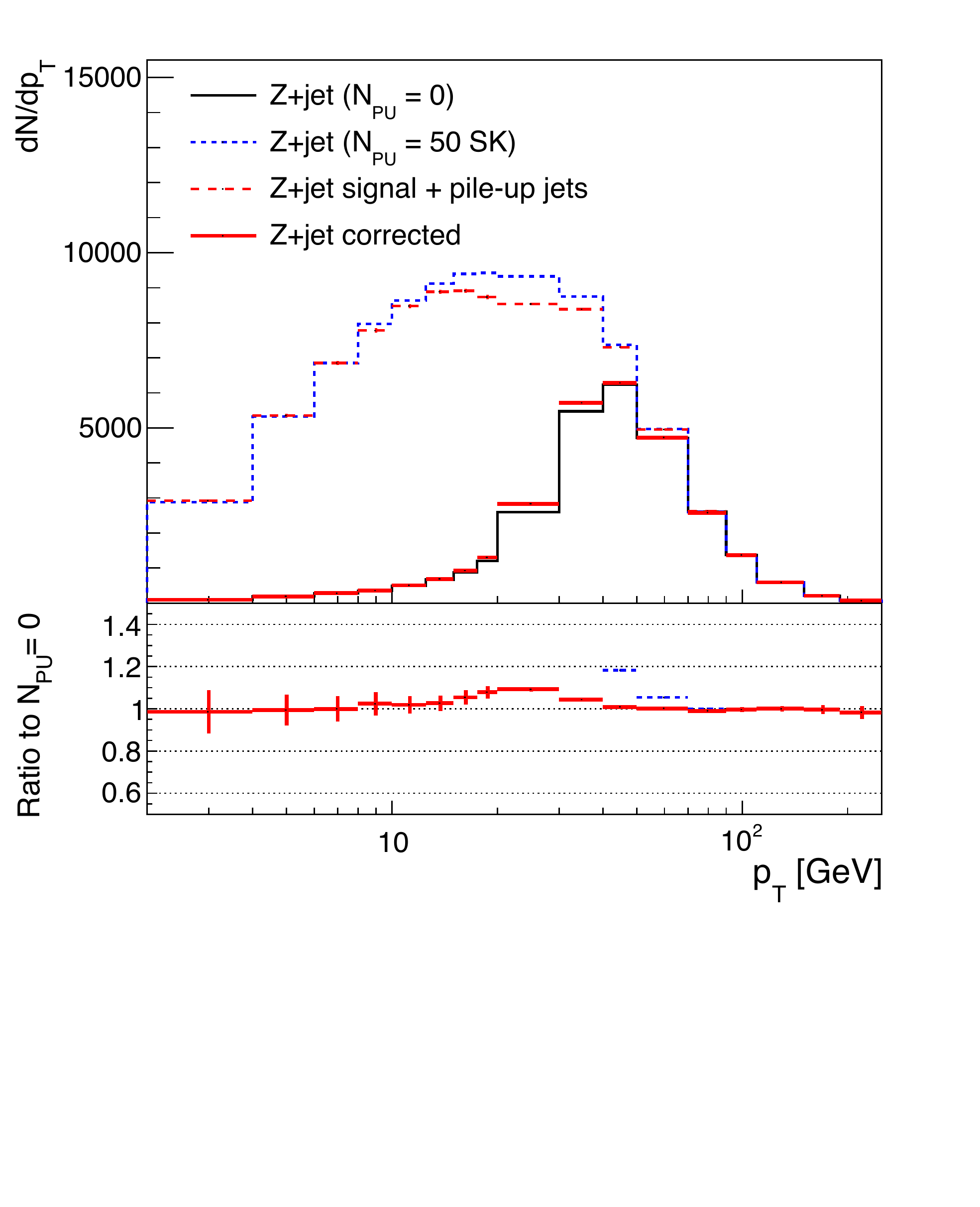}
  \caption{\it Effects of pile-up  (with number of pile-up collisions 
$N_{\rm{PU}} = 50$)   on  
the  $Z$-boson $p_T$ spectrum  in $Z$ + jet production~\protect\cite{Hautmann:2015yya}: 
(left)  application of the pile-up removal method  SoftKiller~\protect\cite{Cacciari:2014gra};    
 (right) application of  the jet mixing 
method~\protect\cite{Hautmann:2015yya}. 
}
\label{fig:plp_fig3}
   \end{center}
\end{figure}

Fig.~\ref{fig:plp_fig3}~\cite{Hautmann:2015yya} illustrates an example of 
$Z$-boson / jet   correlation, showing  
 the  $Z$-boson $p_T$ spectrum  in $Z$ + jet production. On the left  
 one sees the result 
at zero pile-up (solid black curve), 
 the 
result for $N_{\rm{PU}} = 50$ 
pile-up collisions  (dot-dashed black), 
and the result of applying the pile-up removal  method 
SoftKiller~\cite{Cacciari:2014gra} 
(dashed blue); on the right is the 
result of applying the jet mixing 
method~\cite{Hautmann:2015yya} (solid red).  The main observation 
is that while methods designed to take into account jet pedestal and 
soft particles from pile-up work well 
at the level of leading jet spectra, 
 at the level of correlations the effects left over after soft particle removal are non-negligible, and for these  one needs 
additional  methods, of which the jet mixing approach provides an example.   

It is worth noting that the 
proposed jet mixing 
is a statistical method which is data driven, and avoids dependence on 
Monte Carlo modeling. It also has the distinctive feature of not requiring low pile-up runs but rather making use of 
experimental data recorded at high 
pile-up, thus entailing no loss in 
luminosity~\cite{Hautmann:2015rrb}. 

We finally observe that the Run~II  
 Higgs-boson experimental program 
is closely linked to aspects of QCD 
discussed above.  In particular, 
as high statistics is reached in Higgs boson production measurements, a new program of precision studies in gluon fusion at high mass scales becomes 
possible~\cite{Cipriano:2013ooa}, in which the Higgs boson may be 
used  as a color-singlet pointlike source (in the heavy top quark limit) which couples to gluons, and  compared with electroweak sources  coupled to quarks, e.g. Drell-Yan production in a comparable mass range. With this, 
 one will be  able to access for the first time strong-interaction effects such as color correlations,  
 polarized gluons in unpolarized beams,  gluon fusion processes with double spin   flip~\cite{Angeles-Martinez:2015sea,Cipriano:2013ooa}.

\section*{Acknowledgments}

Many thanks  to    the Moriond organizers and staff  for the invitation to 
 a very  interesting   conference.

\end{document}